\documentclass[conference]{IEEEtran}
\IEEEoverridecommandlockouts
\usepackage{amsmath,amssymb,amsfonts}
\usepackage{algorithmic}
\usepackage{graphicx}
\usepackage{textcomp}
\usepackage{xcolor}
\usepackage{empheq}

 \usepackage{graphicx}
 \usepackage{lineno}
 \usepackage{array}
 \usepackage{longtable}
 \usepackage[sort]{natbib}
 \setcitestyle{numbers,square}
 \usepackage{url}
 \usepackage{hyperref} 
 \usepackage{cases}

 \usepackage{amsmath,amsfonts,amssymb}
 \usepackage{bm} 
 \usepackage{float}
 \usepackage{mwe}
 \usepackage{mathtools}
 \usepackage{comment}
 \usepackage{empheq}
 \usepackage{booktabs}
 \usepackage{textcomp}
 \usepackage{algorithm,algorithmic}
 \usepackage[caption=false]{subfig}
 \usepackage{url}
 \usepackage{caption}

 \usepackage{multirow}
 \usepackage{systeme}

 \usepackage{soul}

 \usepackage{tikz}
 \usetikzlibrary{positioning}
 \usetikzlibrary{3d}
 \usetikzlibrary{matrix}
 \usetikzlibrary{decorations.text}
 \usetikzlibrary{spy}

\def\BibTeX{{\rm B\kern-.05em{\sc i\kern-.025em b}\kern-.08em
    T\kern-.1667em\lower.7ex\hbox{E}\kern-.125emX}}
\begin{document}

\title{Combining Weighted Total Variation and Deep Image Prior for natural and medical image restoration via ADMM}

\author{\IEEEauthorblockN{Pasquale Cascarano}
\IEEEauthorblockA{\textit{Department of Mathematics} \\
\textit{University of Bologna}\\
}

\and
\IEEEauthorblockN{Andrea Sebastiani}
\IEEEauthorblockA{\textit{Department of Mathematics} \\
\textit{University of Bologna}\\
}
 \and
 \IEEEauthorblockN{Maria Colomba Comes}
 \IEEEauthorblockA{\textit{Department of Electronic Engineering} \\
 \textit{University of Rome Tor Vergata}\\
}
\and
\IEEEauthorblockN{Giorgia Franchini}
\IEEEauthorblockA{\textit{Department of Mathematics and Computer Science} \\
\textit{University of Ferrara}\\
}

\and
\IEEEauthorblockN{Federica Porta }
\IEEEauthorblockA{\textit{Department of Physics, Informatics and Mathematics} \\
\textit{University of Modena and Reggio Emilia}\\
}
}

\maketitle

\begin{abstract}
In the last decades, unsupervised deep learning based methods have caught researchers attention, since in many real applications, such as medical imaging, collecting a great amount of training examples is not always feasible. Moreover, the construction of a good training set is time consuming and hard because the selected data have to be enough representative for the task. In this paper, we focus on the Deep Image Prior (DIP) framework and we propose to combine it with a space-variant Total Variation regularizer with an automatic estimation of the local regularization parameters. Differently from other existing approaches, we solve the arising minimization problem via the flexible Alternating Direction Method of Multipliers (ADMM). Furthermore, we provide a specific implementation also for the standard isotropic Total Variation. The promising performances of the proposed approach, in terms of PSNR and SSIM values, are addressed through several experiments on simulated as well as real natural and medical corrupted images.
\end{abstract}


\section{Introduction}\label{intro}
The task of image restoration aims to recover a well-looking image, that is clean and sharp, from a blurred and noisy observation. Mathematically, for a given blurred and noisy image $g \in \mathbb{R}^{n}$, the problem can be re-written as an inverse problem of the following form:
\begin{equation}\label{inverse-problem}
    \text{find} \quad u \in \mathbb{R}^{n} \quad s.t. \quad  Hu + \eta = g,
\end{equation}
where $H \in \mathbb{R}^{n \times n}$ is a known operator which models the blur, $\eta \in \mathbb{R}^{n}$ is a realization of the random white Gaussian noise affecting $g$. Problems of the form \eqref{inverse-problem} are well-known to be ill-posed problems \cite{bertero2020introduction}. Therefore, it is impossible to invert the operator $H$ for finding $u$ from \eqref{inverse-problem} due to the lack of stability and/or uniqueness properties. 

In the field of image restoration, different approaches have been proposed in order to provide an estimate $u^{*}$ of the desired solution. The most famous and promising methods can be mainly divided in two categories: regularized reconstruction-based and learning-based methods. 

The regularized reconstruction-based approaches convert the problem into an optimization problem which reads as:
\begin{equation}
    u^{*} \in \underset{u}{\arg\min} \ \dfrac{1}{2}\lVert Hu - g \rVert_{2}^{2} + R(u),
\end{equation}
where the first and the second term are referred to as \textit{fidelity} and \textit{regularization} terms, respectively. The fidelity term models the noise affecting $g$ and, upon zero-mean Gaussian noise assumptions, it is usually defined as an L$_{2}$-norm functional. The regularization term encodes prior information on the solution, such as its sparsity or regularity \cite{karl2005regularization}. 

A popular choice for $R$ is a Total Variation (TV) based functional \cite{rudin1992nonlinear} which, in a discrete setting, is defined as follows: 
\begin{equation}   \label{eq:TV_reg}
\mu \text{TV}(u)=  \mu \sum_{i=1}^{n} \lVert \left( Du \right)_{i} \rVert_{2},
\end{equation}
where $\left( Du \right)_{i} :=\left( \left( D_{h}u\right)_{i},\left(D_{v}u\right)_{i}\right)$  for $i=1 \dots n$, is the discrete gradient of $u$ computed at pixel $i$ and $D_{h}$, $D_{v}$ are the first order finite difference discrete operators along the horizontal and vertical axes, respectively. \\
The positive scalar parameter $\mu$ balances the strength of the regularization. Its choice is crucial to obtain good quality restorations and, in the literature, many methods have been proposed for its selection \cite{katsaggelos1992methods}. Very recently, in order to extend the TV regularizer and to provide a local regularization adapted to the underlying image patterns, a \textit{space-variant} total variation has been proposed \cite{calatroni2020adaptive}. The idea is to weight at any pixel the amount of regularisation by considering the following regularizer:
\begin{equation}   \label{eq:WTV_reg}
\text{WTV}(u)=  \sum_{i=1}^{n} \mu_{i} \lVert \left( Du \right)_{i} \rVert_{2},
\end{equation}
which is more flexible than the standard TV in \eqref{eq:TV_reg}. 

Recently, the learning-based approaches have become popular, in the image restoration field, due to their outstanding performances \cite{mccann2017convolutional}. In particular, the supervised learning-based methods \cite{goodfellow2016deep} make use of Deep Neural Network (DNN) architectures to learn the correlation between the degraded images and their cleaned counterparts from a set of example pairs. In mathematical terms, they attempt to solve the following minimization problem: 
\begin{equation}\label{eq:DL}
    \theta^{*} \in \underset{\theta}{\arg\min} \mathcal{L}(f_{\theta}(G), U), 
\end{equation}
where $f_{\theta}$ is a fixed DNN architecture with weights $\theta$, $\mathcal{L}$ is a fixed loss function and  $\lbrace (G,U) \rbrace$ is a training set of degraded-cleaned example pairs (with $G$ and $U$ we mean the set of  degraded images in input and the cleaned target, respectively). Once \eqref{eq:DL} is solved by means of standard stochastic optimization algorithms, e.g., Adam \cite{kingma2014adam} or Stochastic Gradient Descent (SGD) \cite{bottou2012stochastic}, for a given degraded image $g$, an approximation $u^{*}$ of the desired solution $u$ is obtained as $u^{*}=f_{\theta^{*}}(g)$. \\
The success of this supervised framework is strictly related to the fixed training set of examples. However, in some real applications, such as medical imaging, this is also the main weakness of supervised techniques, since it is practically impossible to collect ground truth data \cite{willemink2020preparing}.

For this reason, unsupervised deep learning based methods have caught researchers attention. One of the most famous unsupervised approach is called Deep Image Prior (DIP), introduced by Ulyanov et al. in \cite{ulyanov2018deep}, which is the combination of the following minimization problem:

\begin{align}\label{eq:Deep-Prior}
 \underset{\theta}{\arg\min} & \dfrac{1}{2}\lVert H f_{\theta}(z) - g \rVert_{2}^{2},
\end{align}
and a regularization by early-stopping procedure. 
In particular, $f_{\theta}$ is a fixed Convolutional Neural Network (CNN) generator whose weights are $\theta$, and $z$ is a random input vector usually sampled from a uniform distribution. The CNN generator is initialized with a set of random weights that are iteratively optimized by means of standard gradient-based algorithms solving \eqref{eq:Deep-Prior}. Then, an approximation $u^{*}$ of the target solution $u$ is computed as $f_{\theta^{*}}(z)$, where $\theta^{*}$ is an early-stopped solution obtained by applying the early-stopping procedure to the involved iterative optimization scheme solving \eqref{eq:Deep-Prior}. In his pioneering work, Ulyanov has empirically shown how the architecture of a Deep CNN is able to represent more easily natural images than random noise, without the need of a fixed set of training examples.

So far, the researchers have mostly worked both on a theoretical analysis of DIP \cite{arridge2019solving,dittmer2020regularization,cheng2019bayesian} and on an improvement of its performances. A widely used approach is to adjust the objective in \eqref{eq:Deep-Prior} in order to constrain the solution to satisfy a given prior.  In particular, in \cite{mataev2019deepred}, the standard DIP objective is combined with regularization by denoising. In \cite{kamilov}, the authors add an explicit anisotropic TV term to \eqref{eq:Deep-Prior}. Furthermore, the idea of combining DIP with TV-based terms has shown good performances when dealing with X-ray images \cite{van2018compressed} and computed tomography reconstructions \cite{Baguer_2020}.


\subsection{Our Contribution}

In this work, we propose to improve the performances of DIP by adding explicit priors to \eqref{eq:Deep-Prior}. Differently from \cite{kamilov}, we consider both the standard isotropic TV and the weighted TV (WTV) defined in \eqref{eq:TV_reg} and \eqref{eq:WTV_reg}, respectively. In particular, we exploit an automatic estimation for all the local regularization parameters when dealing with WTV, thus allowing the implemented algorithm to be less dependent from a good choice of these hyperparameters. More specifically, the regularization parameter $\mu_{i}$ are iteratively updated accordin to the rule proposed in \citep{bortolotti2016uniform} for the Tikhonov regularization. We remark that the WTV functional is more flexible than the standard isotropic TV, since its definition is tailored to better recover the local image patterns.

Furthermore, we make use of the ADMM framework instead of a standard sub-gradient method as in \cite{kamilov}. The ADMM framework is more flexible and due to its modular structure allows us to embed any prior (explicit or implicit) information by simply modifying the regularizer-related substep. 
In the following, we will refer to our proposals as ADMM DIP-TV and ADMM DIP-WTV, to denote the the ADMM implementation when the standard isotropic TV and WTV are used as regularizers, respectively.

The paper is organized as follows: in the second section we introduce our ADMM DIP-TV and ADMM DIP-WTV methods and we show how the resulting ADMM substeps can be efficiently solved. In the third section we present some numerical experiments on synthetic as well as real noisy natural and medical images and we compare the results with the standard DIP \cite{ulyanov2018deep}.

\section{Proposed Method}

The main goal of the proposed ADMM DIP-TV and ADMM DIP-WTV methods is to boost the performances of the DIP framework adding the isotropic TV regularizer in \eqref{eq:TV_reg} and the space-variant WTV in \eqref{eq:WTV_reg}, respectively.
We now describe in details the ADMM DIP-WTV method. Concerning the ADMM DIP-TV method we point out that it is immediate to observe that WTV in \eqref{eq:WTV_reg} reads as the standard TV in \eqref{eq:TV_reg} assuming all $\mu_{i}=\mu,$ $\forall\ i=1 \dots n$. Therefore, it is worth saying that ADMM DIP-TV can be seen as a particular case of ADMM DIP-WTV. 

Let us consider the following minimization problem:
\begin{align}\label{eq:unconstrained:DIP-WTV}
  \underset{\theta}{\arg\min} & \ \dfrac{1}{2} \lVert H f_{\theta}(z) - g \rVert^{2}_{2} + \sum_{i=1}^{N} \mu_{i} \lVert \left( Df_{\theta}(z) \right)_{i} \rVert_{2},
\end{align}
and its constrained counterpart, which reads as:

\begin{align}\label{eq:constrained:DIP-WTV}
   \underset{\theta ,t}{\arg\min} &\dfrac{1}{2} \lVert H f_{\theta}(z) - g \rVert^{2}_{2} + \sum_{i=1}^{N} \mu_{i} \lVert t_{i} \rVert_{2} \\
 \text{s.t.} \quad  & Df_{\theta}(z)=t. \nonumber 
\end{align}

According to the standard DIP framework, we need an iterative process solving \eqref{eq:unconstrained:DIP-WTV}. Therefore,  we attempt to solve the minimization problem \eqref{eq:constrained:DIP-WTV} by means of the ADMM algorithm, which has been recently deeply investigated and applied in a non-convex image restoration framework \cite{hong2016convergence,wang2019global,cascarano2020inverse}. 
The augmented Lagrangian function with respect to the problem \eqref{eq:constrained:DIP-WTV} reads:
\begin{align} \label{eq:lagrangian}
    L(\theta,t, \lambda_{t}) = & \dfrac{1}{2} \lVert H f_{\theta}(z) - g \rVert_{2}^{2} + \sum_{i=1}^{N} \mu_{i} \lVert t_{i} \rVert_{2} \nonumber  \\ + & \dfrac{\beta_{t}}{2}\lVert Df_{\theta}(z) - t \rVert^{2}_{2} + <\lambda_{t}, Df_{\theta}(z) - t >,
\end{align}
where $\beta_{t}$ is a positive scalar, called penalty parameter, $\lambda_{t}$ is the Lagrangian parameter associated with the constraint $Df_{\theta}(z) = t$. According to the ADMM framework, we seek its saddle point by minimizing with respect to the primal variable $\theta$ and $t$, alternatively, and by maximizing with respect to the dual variable $\lambda_{t}$. 
Upon suitable initialization of the variables involved, the $k$-th iteration of the ADMM iterative algorithm reads as follows:

\begin{numcases}{}
\scalebox{1}{$\begin{aligned} \theta^{k+1} \in \underset{\theta}{\arg\min} & \ \dfrac{1}{2} \lVert H f_{\theta}(z) - g \rVert_{2}^{2} + \\ & +  \dfrac{\beta_{t}}{2} \left\lVert D f_{\theta}(z) - t^{k} + \dfrac{\lambda_{t}^{k}}{\beta_{t}} \right\rVert_{2}^{2}  \end{aligned}$} & \label{eq:ADMM_1} \\
\scalebox{1}{$\begin{aligned} t^{k+1}  = \underset{t}{\arg\min} & \ \sum_{i=1}^{n} \mu^{k}_{i} \lVert t_{i} \rVert_{2}^{2} + \\ & + \dfrac{\beta_{t}}{2} \left\lVert t - \left(Df_{\theta_{k+1}}(z) + \dfrac{\lambda_{t}^{k}}{\beta_{t}}\right) \right\rVert^{2}_{2} \end{aligned}$} & \label{eq:ADMM_2} \\
\scalebox{1}{$\lambda_{t}^{k+1} = \lambda_{t}^{k} + \beta_{t}(Df_{\theta^{k+1}}(z) - t^{k+1})$}  & \label{eq:ADMM_3}
\end{numcases}

The first problem \eqref{eq:ADMM_1} is solved inexactly by applying a prefixed number of iterations of a gradient-based method. In particular we use the Adam iterative scheme  \cite{kingma2014adam}. The numerical gradient is computed by means of automatic differentiation provided by Pytorch with respect to the variable $\theta$ \cite{maclaurin2015autograd}.\\ 
We observe that this optimization problem is very similar to the one solved in the classical DIP framework \eqref{eq:Deep-Prior}. In this particular case, we force $Df_{\theta^{k+1}}(z)$ to be close to  $t^{k} - \frac{\lambda^{k}_{t}}{\beta_{t}}$. From a numerical point of view, this squared L$_{2}$-norm term provides a stabilizing and robustifying effect to the DIP minimization. 

The second problem \eqref{eq:ADMM_2} is separable and can be easily solved in a closed form by applying the 2D L$_{2}$-norm proximity operator to the $n$ components of $Df_{\theta^{k+1}}(z) + \frac{\lambda_{t}^{k}}{\beta_{t}}$. In our implementation we chose to vary the regularization parameters $\mu_{i}$ along the iterations. In particular, their formulation is inspired by \cite{bortolotti2016uniform} and reads: 
\begin{equation}
    \mu_{i}^{k}= \dfrac{1}{2n} \dfrac{\lVert H f_{\theta^{k+1}}(z) - g \rVert_{2}^{2}}{\lVert \left(Df_{\theta^{k+1}}(z)\right)_{i} \rVert_{2}}. 
\end{equation}
This entails that the smaller is the gradient magnitude the greater is the regularization provided at pixel $i$. Therefore, we regularize more on constant patches and less on patches with complex texture.

Finally, we remark that concerning the ADMM DIP-TV implementation, we set $\mu_{i}^{k}= \mu$ in problem \eqref{eq:ADMM_2}, where $\mu$ is a fixed positive scalar.

\section{Experiments}

{In this section, we present the results of some numerical experiments aimed at evaluating the effectiveness of the proposed ADMM DIP-TV and ADMM DIP-WTV methods on image denoising problems, that is, on problems of the form  \eqref{inverse-problem} where the operator $H$ is fixed as the identity operator. For all the experiments, we have used the CNN architecture represented in Figure \ref{fig:architecture} which is an adaptation of the U-net architecture proposed in \cite{ulyanov2018deep}.
\begin{figure*}[!ht]
	\centering
	\begin{tikzpicture}[scale=0.6, transform shape]
	
	\tikzset{pics/fake box/.style args={
			#1 with dimensions #2 and #3 and #4 rot #5 text #6}{
			code={
				\draw[gray,ultra thin,fill=#1] (-#2/2,-#3/2) rectangle (#2/2,#3/2);
				\node at (0,0) [rotate=#5, anchor=center] {#6};
			}
			
	}}
    
    \definecolor{max_pool}{HTML}{88D04F};
    \definecolor{skip}{HTML}{1F9DCF};
    \definecolor{upsample}{HTML}{67EB93};
    \definecolor{3conv}{HTML}{97CFF2};
    \definecolor{1conv}{HTML}{D7DC79};
    \definecolor{relu}{HTML}{B2C5F5};
    \definecolor{bn}{HTML}{F6BDAA};
    \definecolor{ui}{HTML}{91E49E};
    \definecolor{di}{HTML}{EAD569};
    
	\draw pic (layer0) at (0,0) {fake box={3conv} with dimensions 0.4 and 4 and 0 rot -90 text {Input}};
	\draw pic (layer1_0) at (1,0) {fake box={di} with dimensions 0.4 and 3.5 and 0 rot -90 text {}};
	\draw pic (layer1_1) at (1.5,0) {fake box={3conv} with dimensions 0.4 and 3.5 and 0 rot -90 text {}};
	\draw pic (layer2_0) at (2.5,0) {fake box={di} with dimensions 0.4 and 3 and 0 rot -90 text {}};
	\draw pic (layer2_1) at (3,0) {fake box={3conv} with dimensions 0.4 and 3 and 0 rot -90 text {}};
	\draw pic (layer3_0) at (4,0) {fake box={di} with dimensions 0.4 and 2.5 and 0 rot -90 text {}};
	\draw pic (layer3_1) at (4.5,0) {fake box={3conv} with dimensions 0.4 and 2.5 and 0 rot -90 text {}};
	\draw pic (layer4_0) at (5.5,0) {fake box={di} with dimensions 0.4 and 2 and 0 rot -90 text {}};
	\draw pic (layer4_1) at (6,0) {fake box={3conv} with dimensions 0.4 and 2 and 0 rot -90 text {}};
	\draw[dashed, gray,thick] (6.5,-2)-- (6.5,1.25);
	\draw pic (layer5_0) at (7,0) {fake box={bn} with dimensions 0.4 and 2 and 0 rot -90 text {}};
	\draw pic (layer5_1) at (7.5,0) {fake box={ui} with dimensions 0.4 and 2 and 0 rot -90 text {}};
	\draw pic (layer6_0) at (8.5,0) {fake box={bn} with dimensions 0.4 and 2.5 and 0 rot -90 text {}};
	\draw pic (layer6_1) at (9,0) {fake box={ui} with dimensions 0.4 and 2.5 and 0 rot -90 text {}};
	\draw pic (layer7_0) at (10,0) {fake box={bn} with dimensions 0.4 and 3 and 0 rot -90 text {}};
	\draw pic (layer7_1) at (10.5,0) {fake box={ui} with dimensions 0.4 and 3 and 0 rot -90 text {}};
	\draw pic (layer8_0) at (11.5,0) {fake box={bn} with dimensions 0.4 and 3.5 and 0 rot -90 text {}};
	\draw pic (layer8_1) at (12,0) {fake box={ui} with dimensions 0.4 and 3.5 and 0 rot -90 text {}};
    \draw pic (layer9) at (13,0) {fake box={3conv} with dimensions 0.4 and 4 and 0 rot -90 text {Output}};
    \draw[thick, skip] (1.5,1.75) -- ++(0,0.5) -| (12,1.75);
    \draw[thick, skip] (3,1.5) -- ++(0,0.4) -| (10.5,1.5);
    \draw[thick, skip] (4.5,1.25) -- ++(0,0.4) -| (9,1.25);
    \draw[thick, skip] (6,1.0) -- ++(0,0.4) -| (7.5,1.0);
    
    \draw[thick,-latex, max_pool] (0.25,0) --++ (0.5,0);
    \draw[thick,-latex, max_pool] (1.75,0) --++ (0.5,0);
    \draw[thick,-latex, max_pool] (3.25,0) --++ (0.5,0);
    \draw[thick,-latex, max_pool] (4.75,0) --++ (0.5,0);
    \draw[thick,-latex, upsample] (7.75,0) --++ (0.5,0);
    \draw[thick,-latex, upsample] (9.25,0) --++ (0.5,0);
    \draw[thick,-latex, upsample] (10.75,0) --++ (0.5,0);
    \draw[thick,-latex, upsample] (12.25,0) --++ (0.5,0);
    
    \node at (6.5,2.05) {Skips Connections};
    \node at (6.5,2.5) {$n_s[i]$};
    \node at (4.1,-1.75) {\LARGE Encoder};
    \node at (9,-1.75) {\LARGE Decoder};
    
    \draw[thick,-latex, skip] (0,-2.45) --++ (0.5,0);
    \node[anchor=west] at (0.5,-2.45) {\small skip};
    \draw[thick,-latex, max_pool] (1.2,-2.45) --++ (0.5,0);
    \node[anchor=west] at (1.7,-2.45) {\small max pool};
    \draw[thick,-latex, upsample] (3.1,-2.45) --++ (0.5,0);
    \node[anchor=west] at (3.6,-2.45) {\small upsample};
    
    \draw pic (legend_0) at (0,-3) {fake box={relu} with dimensions 0.2 and 0.5 and 0 rot -90 text {}};
    \draw pic (legend_0) at (0,-3.75) {fake box={3conv} with dimensions 0.2 and 0.5 and 0 rot -90 text {}};
    \node[anchor=west] at (0.05, -3) {\small Leaky ReLU};
    \node[anchor=west] at (0.05, -3.75) {$3\times 3$ conv.};
    
    \draw pic (legend_0) at (2,-3) {fake box={bn} with dimensions 0.2 and 0.5 and 0 rot -90 text {}};
    \draw pic (legend_0) at (2,-3.75) {fake box={1conv} with dimensions 0.2 and 0.5 and 0 rot -90 text {}};
    \node[anchor=west] at (2.05, -3) {BN};
    \node[anchor=west] at (2.05, -3.75) {$1\times 1$ conv.};
    
    \draw pic (legend_0) at (4,-3) {fake box={ui} with dimensions 0.2 and 0.5 and 0 rot -90 text {}};
    \draw pic (legend_0) at (4,-3.75) {fake box={di} with dimensions 0.2 and 0.5 and 0 rot -90 text {}};
    \node[anchor=west] at (4.05, -3) {$u[i]$};
    \node[anchor=west] at (4.05, -3.75) {$d[i]$};
    
    \draw[gray,ultra thin,fill=di] (6.55,-4) rectangle (8.05,-3.1);
    \draw[-latex, gray] (6.4,-3.2) --++ (2,0);
    \draw pic  at (6.7,-3.5) {fake box={relu} with dimensions 0.2 and 0.75 and 0 rot -90 text {}};
    \draw pic  at (7,-3.5) {fake box={bn} with dimensions 0.2 and 0.75 and 0 rot -90 text {}};
    \draw pic  at (7.3,-3.5) {fake box={3conv} with dimensions 0.2 and 0.75 and 0 rot -90 text {}};
    \draw pic  at (7.6,-3.5) {fake box={bn} with dimensions 0.2 and 0.75 and 0 rot -90 text {}};
    \draw pic  at (7.9,-3.5) {fake box={relu} with dimensions 0.2 and 0.75 and 0 rot -90 text {}};
    
    \draw[gray,ultra thin,fill=ui] (8.75,-4) rectangle (10.85,-3.1);
    \draw[-latex, gray] (8.6,-3.2) --++ (2.55,0);
    \draw pic  at (8.9,-3.5) {fake box={1conv} with dimensions 0.2 and 0.75 and 0 rot -90 text {}};
    \draw pic  at (9.2,-3.5) {fake box={bn} with dimensions 0.2 and 0.75 and 0 rot -90 text {}};
    \draw pic  at (9.5,-3.5) {fake box={relu} with dimensions 0.2 and 0.75 and 0 rot -90 text {}};
    \draw pic  at (10.1,-3.5) {fake box={3conv} with dimensions 0.2 and 0.75 and 0 rot -90 text {}};
    \draw pic  at (10.4,-3.5) {fake box={bn} with dimensions 0.2 and 0.75 and 0 rot -90 text {}};
    \draw pic  at (10.7,-3.5) {fake box={relu} with dimensions 0.2 and 0.75 and 0 rot -90 text {}};
    
    \node[] at (11.95, -2.9) {$n_s[i]$};
    \draw[gray,ultra thin,fill=skip] (11.5,-4) rectangle (12.4,-3.1);
    \draw[-latex, gray] (11.35,-3.2) --++ (1.35,0);
    \draw pic  at (11.65,-3.5) {fake box={1conv} with dimensions 0.2 and 0.75 and 0 rot -90 text {}};
    \draw pic  at (11.95,-3.5) {fake box={bn} with dimensions 0.2 and 0.75 and 0 rot -90 text {}};
    \draw pic  at (12.25,-3.5) {fake box={relu} with dimensions 0.2 and 0.75 and 0 rot -90 text {}};
	\end{tikzpicture}
	\caption{The CNN architecture used in our methods.}
	\label{fig:architecture}
\end{figure*}
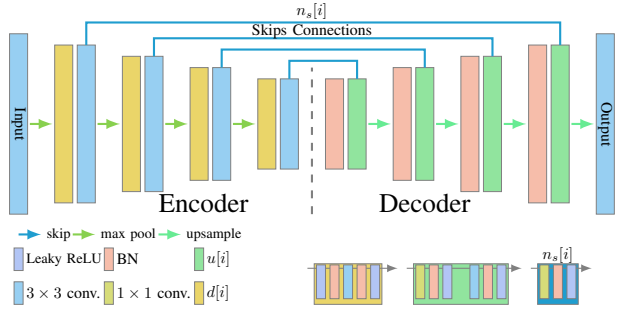
We consider a dataset of four images composed by two natural RGB  images, one synthetic grayscale image and one real medical chest CT image of a patient affected by COVID-19 \cite{yan2020automatic}. Both the ground truths related to the first three images and the acquired CT image are shown in Figure \ref{fig:imageset}.}


\begin{figure}[!ht]
    \centering
    \begin{tabular}{cc}
         \includegraphics[width=0.15\textwidth, height=0.15\textwidth]{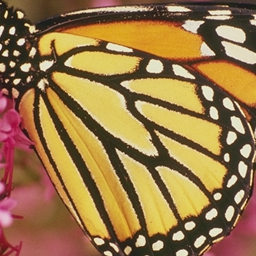} & \includegraphics[height=0.15\textwidth]{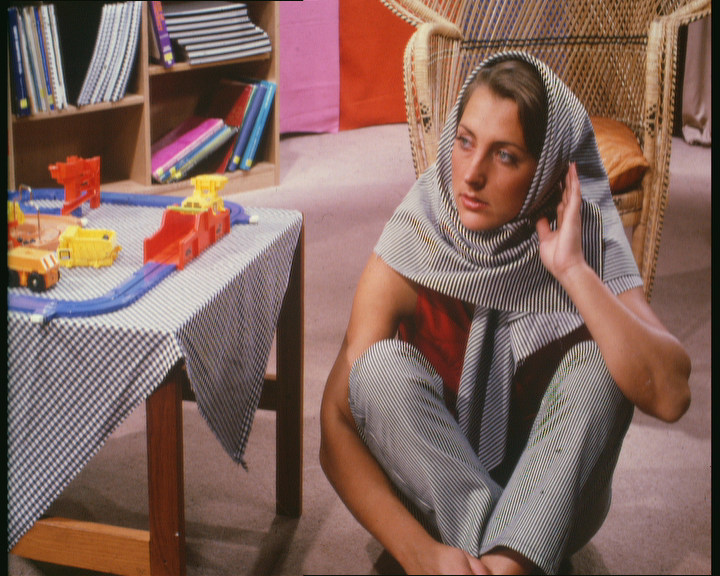} \\
         \textit{Butterfly} ground truth & \textit{Barbara} ground truth\\
         \includegraphics[width=0.15\textwidth, height=0.15\textwidth]{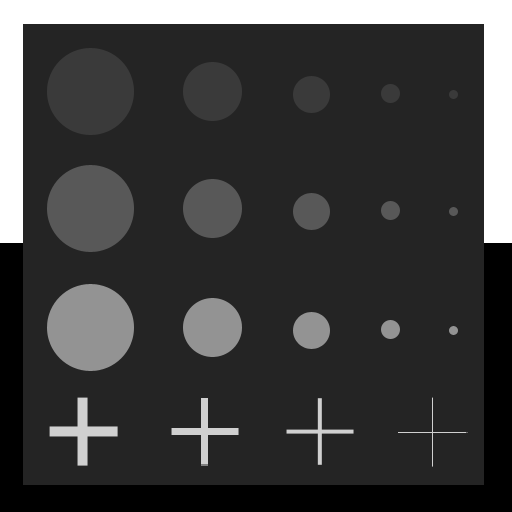} & \includegraphics[width=0.15\textwidth, height=0.15\textwidth]{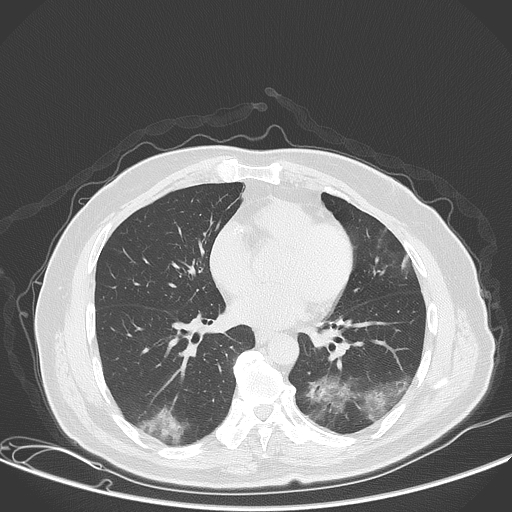}\\
         \textit{Tomo} ground truth & CT acquired data\\
    \end{tabular}
    \caption{The images dataset employed in the numerical experiments.}
    \label{fig:imageset}
\end{figure}

{The starting degraded images corresponding to the natural and the synthetic data are created by applying the image formation model \eqref{inverse-problem} to the related ground truths. The codes and the images used for these numerical experiments are available online\footnote{\url{https://github.com/sedaboni/ADMM-DIPTV}}}.

As a first test, we take into account the denoising task of the geometric grayscale \textit{tomo} image. 
This image has some low contrast and high contrast patches with big and small objects. We corrupt the ground truth by adding a white Gaussian noise with standard deviation equals to 30. The starting noisy image is reported in the first panel of Figure \ref{fig:tomo} with two different close-ups highlighting small and big details. Since the \textit{tomo} image is an example of a piecewise constant image, it is well known that a good reconstruction can be obtained by employing, as a regularization term, the TV functional. For this reason, this test problem fits well as a first benchmark to highlight the benefits of applying a TV-based constraint to the DIP algorithm. In the second and third panels of Figure \ref{fig:tomo} we report the reconstructed images provided by DIP and ADMM DIP-TV. For the ADMM DIP-TV method we perform 5 Adam iterations at each stage of the iterative procedure to update $\theta^{k+1}$.
The image obtained by applying the DIP approach shows some issues in reconstructing the low contrast patches in the image: the edges are not sharp and look out of focus, and the small details are not perfectly retrieved. Moreover, the two close-ups
reveal the presence of artifacts over the edges and the noise seems to not be perfectly removed. The addition of TV to the standard DIP framework seems to solve all the aforementioned issues. In the reconstruction provided by the ADMM DIP-TV scheme, the low contrast small circles are perfectly retrieved while the higher contrast details show sharp edges.

\begin{figure}
	\centering
	\includegraphics[width=.50\textwidth]{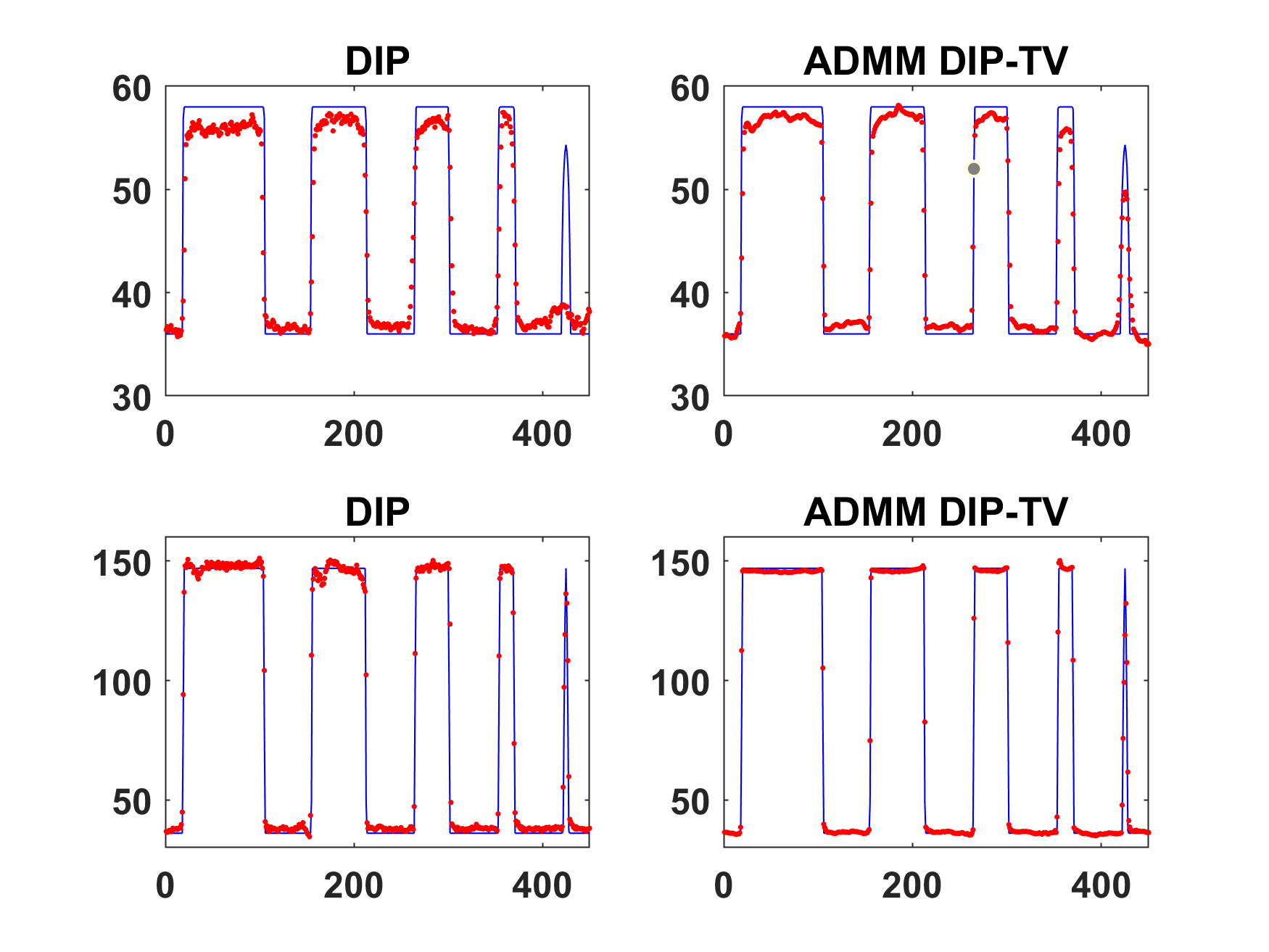}
	\caption{{Reconstructed line profiles (red) provided by DIP (first column) and ADMM DIP-TV (second column) over the ground truth line profiles (blue). The first row refers to a low contrast line profile (90-$th$), while the second row to an high contrast line profile (370-$th$). }}
	\label{fig:lineprofile}
\end{figure}

In order to analyse in details the results on this synthetic image over low and high contrast details, we represent in Figure \ref{fig:lineprofile} the line profiles of rows $90$ (low contrast line profile) and $370$ (high contrast line profile). The first and the second rows of Figure \ref{fig:lineprofile} refer to the low and high contrast line profiles, respectively. We show by red lines the reconstructed line profiles provided by DIP and ADMM DIP-TV, all of them superimposed on to the ground truth line profiles depicted in blue. It is evident that the standard DIP is insufficient to get rid of the noise since both the low and high contrast line profiles look swinging. The low contrast line profile obtained by DIP misses the small low contrast peak, which corresponds to the small low contrast circle in the solution by DIP reported in Figure \ref{fig:tomo}. As expected, this simple test shows how a TV-based method is more efficient in well recovering piecewise constant images with respect to the standard DIP. However, we remark that the good ADMM DIP-TV reconstruction have been obtained by properly tuning the regularization parameter $\mu$ in \eqref{eq:TV_reg}. Not suitable choices of $\mu$ can lead to bad reconstructed images. This well known strong dependency of the results from the regularization parameter can be overcome by using the WTV regularizer, since its weights are adaptively estimated during the iterative process. For the other test problems considered in this section, we compare the standard DIP method with both the ADMM DIP-TV and ADMM DIP-WTV algorithms.

\input{fig2}

\input{fig6}

\begin{figure}
	\centering
	\includegraphics[width=.50\textwidth]{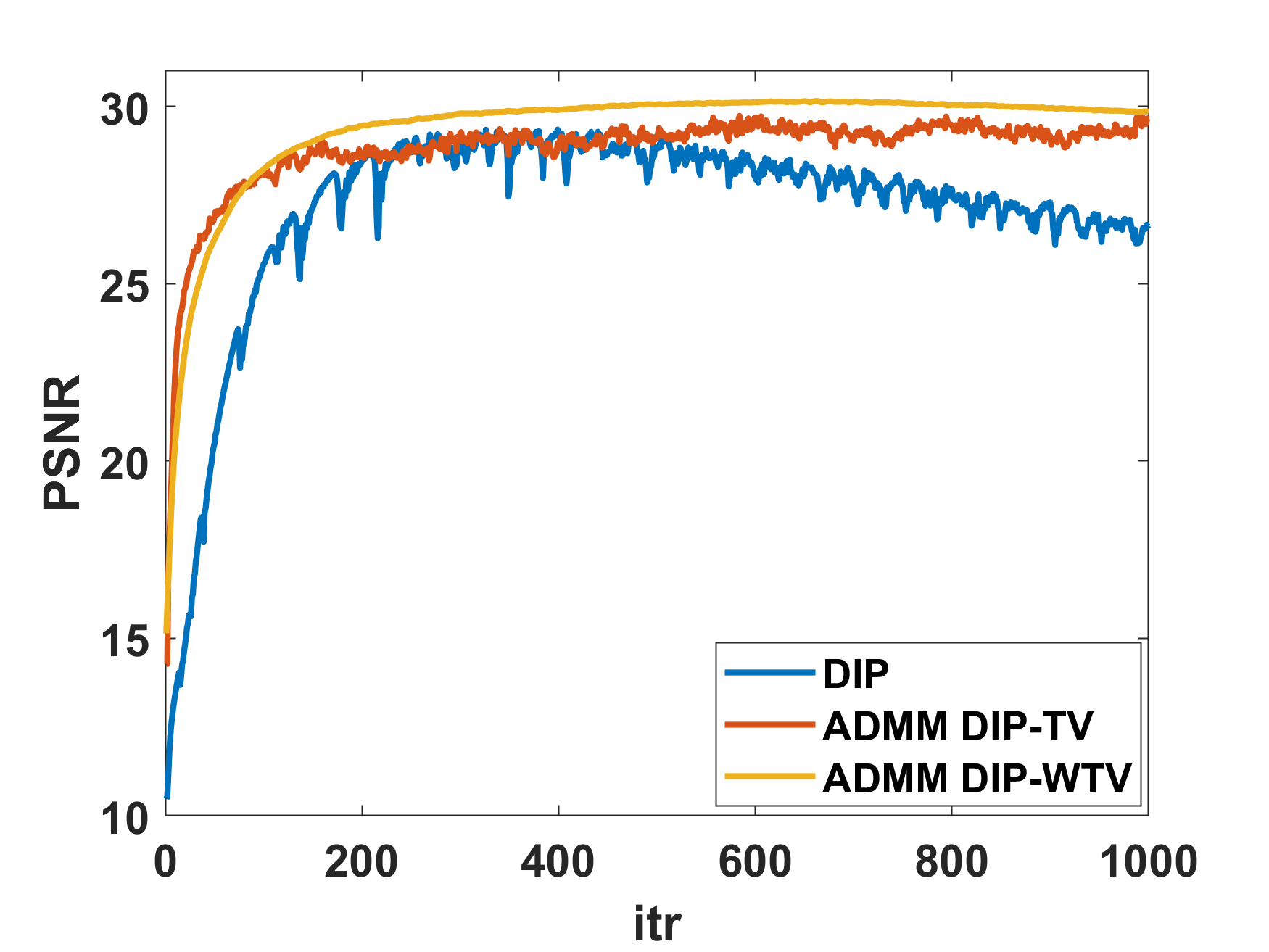}
	\caption{The PSNR values achieved by DIP, ADMM DIP-TV and ADMM DIP-WTV along the outer iterations for the \textit{butterfly} test problem.}
	\label{fig:plot}
\end{figure}

{We firstly report the results on the \textit{butterfly} test problem, whose ground truth has been corrupted by Gaussian noise with standard deviation equals to 20. The simulated acquisition is depicted in the first panel of Figure \ref{fig:6}. Moreover, the other three panels of Figure \ref{fig:6} show the best restored images, in terms of PSNR, provided by DIP, ADMM DIP-TV and ADMM DIP-WTV. For ADMM DIP-TV and ADMM DIP-WTV, 50 Adam iterations have been performed to solve the sub-problem in the primal variable. It is quite evident how both the reconstructions obtained by employing ADMM DIP-TV and ADMM DIP-WTV are sharper and more faithful to the original image than that achieved by DIP. However, in the two close-ups it is possible to appreciate the effect of the WTV in better removing the noise presence and efficiently recovering the image discontinuites. As reported in the caption of Figure \ref{fig:6}, by adding the TV-based regularizers we observe that the PSNR metric increases. In particular ADMM DIP-WTV outperforms ADMM DIP-TV in terms of both PSNR and SSIM. 

As mentioned before, the reconstructed images, offered in Figure \ref{fig:6}, are related to the best PSNR values achieved by the different methods during the iterative process. Of course, such values are not known when we consider imaging problems with real data. For this reason we analyze the PSNR trend for the three approaches under consideration. Particularly, we investigate how the addition of the handcrafted TV acts on the typical semiconvergence behaviour of the DIP framework \cite{ulyanov2018deep}. In Figure \ref{fig:plot} we report the PSNR values along the outer iterations. 
We observe that 
DIP performances rapidly degrades while both ADMM DIP-TV and ADMM DIP-WTV show a more stable behaviour. In particular, the PSNR curve corresponding to ADMM DIP-WTV not only does not semiconverge but also does not present significant oscillations. The ADMM DIP-WTV algorithm seems to be the most robust one in terms of good restored images. To summarize, in real applications, since the best PSNR is not available and, hence, it is not possible to know when the iterative process should be stopped, it is preferable to use the TV-based ADMM approaches. However, even if both ADMM DIP-TV and ADMM DIP-WTV manage to prevent the semiconvergence effect, we remark again that for the ADMM DIP-TV method a pre-processing step is required to select a good regularization parameter while the ADMM DIP-WTV algorithm is not influenced by the choice of the initial value of such parameter.}

 Concerning the \textit{Barbara} test problem, the corrupted image (reported in the first panel of Figure \ref{fig:7}) has been generated by adding to the original one Gaussian noise with standard deviation equals to 30. ADMM DIP-TV and ADMM DIP-WTV solve problem \eqref{eq:ADMM_1} with 50 Adam iterations. Figure \ref{fig:7} show the reconstructed images obtained by the three approaches at the best PSNR achieved. Similar considerations to those made for the results of the \textit{butterfly} test problem also hold in this case. The benefits in employing ADMM DIP-TV and ADMM DIP-WTV are clear. Moreover, the two close-usp of Figure \ref{fig:7} show that ADMM DIP-WTV also outperforms ADMM DIP-TV in recovering the details of the dress texture and the edges of the bookcase on the background. The different PSNR behaviour provided by the three methods observed in Figure \ref{fig:plot} for the \textit{butterfly} test problem can be also observed for the \textit{Barbara} test problem.
\input{fig7}

As last problem, we consider a real chest CT medical image representing a section of two lungs of a patient affected by COVID-19 \cite{yan2020automatic}. The first column of Figure \ref{fig:covid_ct} reports the acquired data together with the close-ups of two details (inflammation zones) in the lungs backside where are visible the effects of the interstitial pneumonia caused by COVID-19 disease. From these panels the standard artifacts related to the discrete angles sampling typical of the CT application are clearly visible. In the second column of Figure \ref{fig:covid_ct} we show the best denoised images provided by the original DIP approach. In order to obtain the best recovered image, we printed the reconstructed images at each DIP iteration and we stopped the algorithm when we visually found a good one, before DIP started to re-introduce the noisy artifacts in the reconstructions. A total of 300 iterations have been carried out for the DIP method to achieve the images in Figure \ref{fig:covid_ct}. The third and the fourth column of Figure  \ref{fig:covid_ct} report the best denoised images attained by ADMM DIP-TV and ADMM DIP-WTV, respectively. The best images have been obtained as explained before for the DIP case, after 10 iterations. However, we remark that, for each of the outer iterations, 500 Adam iterations have been performed to inexactly solve the sub-problem for $\theta$. We also observe that by increasing the outer iterations, both ADMM DIP-TV and ADMM DIP-WTV do not recover the CT artifacts as strongly as the DIP method does, but the recovered images are quite stable in terms of details detection and noise reduction. This aspect allows easier stopping techniques for both the ADMM approaches than those for the DIP one. From Figure \ref{fig:covid_ct}, it is possible to conclude that the best recovered images have been provided by the ADMM DIP-WTV method: the edges are sharper and all finer structures are sufficiently well reconstructed. The yellow arrows in the close-ups highlight the inflammation details, alveoli and bronchioles. It is evident that the restoration provided by ADMM DIP-WTV is more reliable than the ones provided by DIP and ADMM DIP-TV. 
\input{fig5}

\section{Conclusion}
In this paper we have presented a new algorithm which extends the classical DIP framework by adding either the standard isotropic TV or the space-variant TV. The latter admits an automatic strategy for the estimation of the local regularization parameters, thus resulting more flexible with respect to the TV functional and providing more reliable restorations.
The arising optimization problems have been solved in a flexible ADMM framework. The usage of the ADMM splitting ensures stability to the algorithm as the noise increases and allows to add different priors to the standard DIP framework. 
Numerical experiments have shown that our approach reaches better performances than the standard DIP for the denoising of both synthetic and real natural and medical images.

\section*{Acknowledgment}

This work has been partially supported by GNCS-INDAM grant 2020 \textit{Ottimizzazione per l’apprendimento automatico e apprendimento automatico per l’ottimizzazione} and POR-FSE 2014-2020 funds of Emilia-Romagna region (Deliberazione di Giunta Regionale n. 255- 30/03/2020).

\end{document}